\documentclass[prl,twocolumn,showpacs,showkeys,preprintnumbers,amsmath,amssymb,superscriptaddress]{revtex4}

\usepackage{graphicx}
\usepackage{dcolumn}
\usepackage{bm}
\usepackage{color}
\usepackage{psfrag}

\usepackage{amsmath}
\usepackage{amssymb}
\newcommand{\abbrev}[1]{\textsc{#1}}
\newcommand{\inlinesup}[1]{{\scriptscriptstyle #1}}

\begin{document}

\title{Properties of Vacancy Formation in \textsuperscript{4}He hcp Crystals at Zero Temperature \\and Fixed Pressure}

\author{Y.\ Lutsyshyn}
\email{yaroslav.lutsyshyn@upc.edu}
\affiliation{
Departament de F\'{i}sica i Enginyeria Nuclear, Universitat Polit\`{e}cnica de Catalunya, 
Campus Nord B4-B5, E-08034, Barcelona, Spain}
\author{C.\ Cazorla}
\affiliation{
Department of Chemistry, University College London, London WC1H 0AJ, UK
}
\author{G.\ E.\ Astrakharchik}
\author{J.\ Boronat}
\affiliation{
Departament de F\'{i}sica i Enginyeria Nuclear, Universitat Polit\`{e}cnica de Catalunya, 
Campus Nord B4-B5, E-08034, Barcelona, Spain}

\date[Compiled: ]{\today}

\begin{abstract}
Equation of state of \textsuperscript{4}He hcp crystals with 
vacancies is determined at zero temperature using the diffusion Monte Carlo technique, 
an exact ground state zero-temperature method.
This allows us to extract the formation enthalpy 
and isobaric formation energy of a single vacancy in otherwise perfect helium solid.
Results were obtained for pressures up to 160~bar.
The isobaric formation energy is found to reach a minimum  
near 57 bar where it is equal to $10.5\pm 1.2$~K.
At the same pressure, the vacancy formation volume exhibits a maximum and reaches the volume of the unit cell.
This pressure coincides with the pressure interval over which a peak 
in the supersolid fraction of \textsuperscript{4}He was observed in a recent experiment.
\end{abstract}

\maketitle

Properties of solid \textsuperscript{4}He have regained interest since
the discovery that it exhibits nonclassical moment of inertia \cite{Chan2004} and several accompanying phenomena 
(for a review, see Refs~\cite{Prokof'ev2007,Reatto2008,Balibar2008}).
Recently, Kim and Chan studied samples grown at fixed pressure  \cite{Chan2006}.
They found that the supersolid fraction changed threefold over the studied pressure range (up to 135~bar)
and had a distinct maximum at pressures near 55 bar, which is well above melting pressure.
These results call for a study of possible supersolidity mechanisms under isobaric conditions.

Crystal defects are believed to be indispensable in the mechanisms behind supersolidity  
\cite{Prokof'ev2007,Prokof'ev2005SupersolidStateOfMatter}.
In particular, special attention has been paid in the past to vacancy defects.
Unbound vacancies, whether intrinsic to solid helium or introduced by experimental conditions, 
are likely to cause supersolidity \cite{Balibar2008,Boronat2009NJP}.
Experimental results for energy of 
vacancy formation in pure \textsuperscript{4}He are available from
the work of Fraass et al.~\cite{Simmons1989}, with additional 
data and analysis by Blackburn and 
colleagues~\cite{Blackburn2007AbsenceOfALowTemperatureAnomalyInTheDebyeWallerFactorOfSolidHe4DownTo140mK}.
Numerous theoretical works describe vacancies in 
solid helium 
\cite{Ceperley2004-RingExchangesAndTheSupersolidPhaseOfHe4,
Reatto1997,Chester1999,Reatto2004,
Boninsegni2006-FateOfVacancyInducedSupersolidityInHe4,
Ceperley2008-PathIntegralCalculationsOfVacanciesInSolidHelium,
Boninsegni2008-LocalStressAndSuperfluidProperties,Reatto2009,Boronat2009NJP,
Vitiello2009-ZeroPointDivacancyConcentrationInTheShadowWaveFunctionModelForSolidHe4}.
Results for vacancy formation energy are available from 
finite-temperature calculations using path integral Monte Carlo methods
\cite{Boninsegni2006-FateOfVacancyInducedSupersolidityInHe4,
Ceperley2008-PathIntegralCalculationsOfVacanciesInSolidHelium}
and from ground-state calculations 
with variational \cite{Reatto1997,Chester1999,Reatto2004,Reatto2009}, 
shadow path integral ground state \cite{Reatto2009} 
and diffusion \cite{Boronat2009NJP} Monte Carlo methods.
These calculations agree that individual vacancies cost too much energy 
to exist in the ground state of solid helium.
However, properties of isobaric 
vacancy formation in \textsuperscript{4}He have not yet been reported.
Additionally, several important vacancy properties, such as vacancy volume,
are only sensible if considered at fixed pressure.
To address these concerns, we performed a calculation of the equation of state 
of solid \textsuperscript{4}He with the number of atoms incommensurate
with the number of sites of an hcp lattice filling the volume.
Incommensurate in this sense, the solid remained crystalline and had 
one vacancy defect per (periodic) simulation volume.
Obtaining such an equation of state allowed us to extract the dependence of vacancy properties
computed at both fixed pressure and fixed density conditions.
Importantly, calculated formation energy at fixed pressure is considerably smaller 
than the formation energy at fixed density. 
It also turned out to exhibit a minimum around the same pressure 
where a peak in the supersolid response was observed experimentally.

The most important thermodynamic quantity characterizing the presence of vacancies in fixed-pressure systems
is the Gibbs free energy \cite{AshcroftMermin}, which at zero temperature reduces to enthalpy.
Consider the enthalpy $H(N,0,P)$ of a commensurate system consisting of $N$ particles 
at pressure $P$, and $H(N-1,1,P)$, the enthalpy
of an incommensurate system with $N-1$ particles and one vacancy at the same pressure $P$.
If $h$ is the enthalpy per particle and $h_{\text{vac}}$ the vacancy formation enthalpy, then
$H(N,0,P)=N h(P)$ and $H(N-1,1,P)=(N-1) h(P) + h_\text{vac}(P)$.
The vacancy formation enthalpy in a solid can therefore be expressed~\cite{Gulseren2003} as 
\begin{eqnarray}
h_\text{vac}(P)&=&H(N-1,1,P)-\frac{N-1}{N}\:H(N,0,P) \nonumber\\
&=&E(N-1,1,P)-\frac{N-1}{N}\:E(N,0,P) \nonumber\\
& & {} \quad \quad +(N-1)\:P\left[\frac{1}{\rho_\text{inc}}-\frac{1}{\rho_\text{com}}\right]\text{,}
\label{enthformexpr}
\end{eqnarray}
where $\rho_\text{inc}$ and $\rho_\text{com}$ are the densities of 
the incommensurate and commensurate systems at pressure $P$.
Pressure can obtained from the derivative of the energy with respect to the density $\rho$ as
\begin{equation}
P=\rho^2\frac{\partial E/N}{\partial \rho}\text{,}
\label{pressure}
\end{equation}
thus allowing us to obtain $E(P)$ and $\rho(P)$ from the $E(\rho)$ dependence available from the calculations.

The isobaric vacancy formation energy $\Delta E_P$ is the energy cost
of moving a single atom away from a lattice site, while keeping
the pressure fixed,
\begin{equation}
\Delta E_P=E(N-1,1,P)-\frac{N-1}{N}E(N,0,P)\text{.}
\label{energyatpressure}
\end{equation}
$\Delta E_P$ determines the formation enthalpy, with the addition of the work necessary to 
free the vacancy volume $\Omega_\text{vac}$, via $h_\text{vac}=\Delta E_P+P \Omega_\text{vac}$. 
The incommensurate system has an excess of volume with regard to its number of atoms. 
It can be written as $V(N-1,1,P)=(N-1) v+\Omega_\text{vac}$ where $v$ is 
the volume normally occupied per particle, $V(N,0,P)=Nv$.
Hence the vacancy volume can be obtained from
\begin{equation}
\Omega_\text{vac}=V(N-1,1,P)-\frac{N-1}{N}V(N,0,P)\text{.}
\label{vacancyvolume}
\end{equation}

In previously published numerical simulations, the formation energy was calculated at \emph{fixed density} as
\begin{equation}
\Delta E_\rho=E(N-1,1,\rho)-\frac{N-1}{N}E(N,0,\rho)\text{,}
\label{energyatdensity}
\end{equation}
which compares two systems with volume adjusted to provide equal densities.
While $\Delta E_\rho$ matches $\Delta E_P$ at \emph{zero} pressure \cite{Gillan1989}, they generally differ
at non-zero pressures necessary to solidify helium. In fact, $\Delta E_\rho$ slowly approaches $h_\text{vac}$ 
in the thermodynamic limit: $h_\text{vac} - \Delta E_\rho \propto 1/N$   \cite{SuplementaryMaterialToActPaper}.
Attempting to obtain $h_\text{vac}$ from $\Delta E_\rho$ 
is thus subject to finite-size effects even if $h_\text{vac}$ itself converges quickly.

Calculations were made with diffusion Monte Carlo (\abbrev{dmc}), a statistically exact ground-state
method.  DMC has proven indispensable in understanding properties of superfluid \textsuperscript{4}He,
and yields excellent results for the properties of solid \textsuperscript{4}He
\cite{Boronat2005-QuantumMonteCarloSimulationOfOverpressurizedLiquidHe4,Boronat2009NJP}, 
in particular for the equation of state.
DMC projects the excited states from the initial conditions $\phi_\abbrev{i}$ by advancing 
through imaginary time $\tau$ a function
$f=\phi_\abbrev{g}\exp[-(H-E_{0})\tau]\phi_\abbrev{i}$, where $H$ is the Hamiltonian, $E_{0}$ 
is a reference energy, and $\phi_\abbrev{g}$ is an importance sampling function, 
also called guiding wavefunction. 
For details of the method, see Refs~\cite{Chin1990,Boronat1994,LesterBook1994}.
Interactions were modelled with the Aziz \abbrev{hfd-b(he)} potential \cite{AzizII}. 
While this potential is known to produce a slight systematic bias in energy of under 70~mK, it 
allows for a good reproduction of the equation of state $P(\rho)$ 
\cite{Boronat1994,Boronat2005-QuantumMonteCarloSimulationOfOverpressurizedLiquidHe4}.

\begin{figure}
\psfrag{densitynnc}[][c]{$\rho$, [nm$^\inlinesup{-3}$]}
\psfrag{eperparticle}[][c]{$e$, [K]}
\psfrag{densnnc}[][c]{$\rho$, [nm$^\inlinesup{-3}$]}
\psfrag{presbar}[][c]{$P$, [bar]}
\psfrag{errormK}[][c]{$\sigma_e$, [mK]}
\psfrag{melting}{melting}
\includegraphics[angle=0,width=\columnwidth]{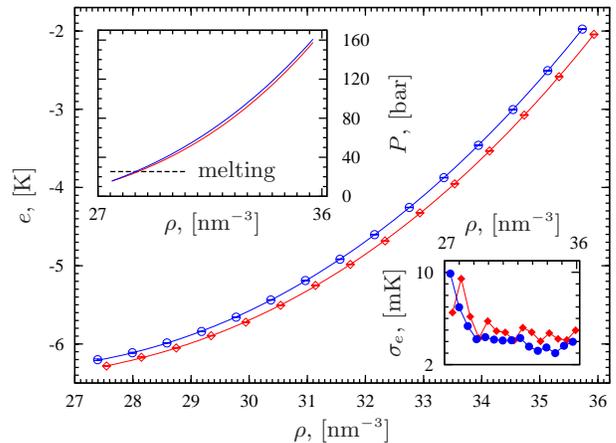}
\caption{\label{fig:energy} (Color online) Energy per particle of the commensurate system ({\color{red}$\lozenge$})
and the incommensurate system with one of the 180 sites being vacant ({\color{blue} $\circ$}).
Error bars are smaller than the symbols and are shown within them. 
Solid lines in the main plot area are third-degree polynomial fits as described in the text.
Lower inset separately shows statistical errors with solid symbols.
Top inset shows pressures for both systems (pressure of the incommensurate solid is higher). Dashed line
marks melting pressure, below which both systems are metastable but remain crystalline.}
\end{figure}

A good-quality guiding wavefunction $\phi_\abbrev{g}$ is necessary to efficiently sample the
energy from the ground state itself.
Such a wavefunction has to take into account two-body correlations 
between atoms, to provide lattice symmetry,
to allow for hopping between lattice sites, and finally it needs to satisfy the Bose symmetry. 
We used a symmetrized form of the Nosanow--Jastrow  \cite{Nosanow1964} wavefunction
that was recently developed for quantum solids~\cite{Boronat2009NJP} 
and has the form 
\begin{equation}
\phi_\abbrev{snj}=\left(\prod_{i<j}^{N_{\text{p}}} f\left(\left|\bm{r}_i-\bm{r}_j\right|\right) \right)
\left( \prod_{k}^{N_{\text{s}  \makebox[0cm]{\phantom{$\scriptscriptstyle p$}}     }} 
\sum_i^{N_\text{p}} g\left(\left|\bm{r}_i-\bm{l}_k\right|\right)\right)\text{,}
\label{snj}
\end{equation}
where $N_\text{p}$ and $N_\text{s}$ are respectively the number of 
atoms and lattice sites, $f(r)$ is the pair correlation
function and $g(r)$ is a function that localizes atoms to the lattice sites.
$|\bm{r}_i-\bm{l}_k|$ denotes distance from an atom with index $i$ to the lattice site $k$.
$\phi_\abbrev{snj}$ is both Bose-symmetric and provides an excellent spatial order~\cite{Boronat2009NJP}.
We used $g(r)=\exp[-1/2 \gamma r^\inlinesup{2}]$ 
and the McMillan form $f(r)=\exp[-1/2 \: (b/r)^\inlinesup{5}]$.
The parameters were obtained from variational optimization of the Nosanow--Jastrow wavefunction, resulting in
$b=1.12\sigma$, where $\sigma=2.556\text{\AA}$, 
and $\gamma=(-6.120+27.69\rho\sigma^3)/\sqrt{\sigma}$. 
Stability was observed for changes of timestep, population size and wavefunction parameters
over the entire range of densities.

\begin{figure}
\psfrag{densitynnc}[][c]{$\rho$, [nm$^\inlinesup{-3}$]}
\psfrag{acterhoK}[][c]{$\Delta E_\rho$, [K]}
\includegraphics[angle=0,width=\columnwidth]{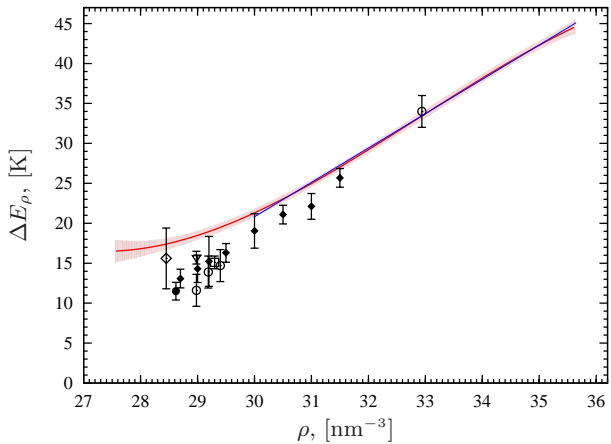}
\caption{\label{fig:energyatfixeddensity} (Color online) Formation energy of a vacancy at fixed density, as defined 
by Eq.~(\ref{energyatdensity}). Energy is shown with the solid line while the area within one standard deviation
is shaded. 
The straight line shows the fit at high densities.
Symbols show numerical data from 
Refs~
\cite{Reatto1997}$~(\circ)$, \cite{Chester1999}$~(\lozenge)$,
\cite{Reatto2004}$~(\triangledown)$, \cite{Reatto2009}$~(\square)$,
\cite{Ceperley2008-PathIntegralCalculationsOfVacanciesInSolidHelium}$~(\bullet)$
and \cite{Boninsegni2006-FateOfVacancyInducedSupersolidityInHe4}$~(\blacklozenge)$.
}
\end{figure}

The commensurate solid was simulated by considering a periodic system 
with 180 atoms in a volume containing 180 hcp lattice sites. 
This number allows for a nearly 
cubic geometry, with ratio of sides of the simulation box equal to $1.06\!:\!1.02\!:\!1$.
The incommensurate crystal was simulated by using one less atom while keeping the lattice intact. 
Size effects resulting from finite simulation box size 
were compensated by separate variational calculations involving up to 1440 atoms with
the methodology of Ref.~\cite{Boronat2008} 
but with the inclusion of $(1/N)^\inlinesup{2}$  terms.
Vacancy images created by periodic boundary conditions 
are sufficiently separated to neglect their interaction \cite{Mahan2006VacancyVacancyInteractionInSupersolids}.

The results for the equation of state of both commensurate and incommensurate 
systems are presented in Fig.~\ref{fig:energy}. 
The resulting density dependence of energy per particle $e(\rho)$
could be fitted accurately with third-degree polynomials,
which in this case are equivalent to the form 
\begin{equation}
e=e_0+b\,(\rho/\rho_0-1)^2+c\,(\rho/\rho_0-1)^3
\label{eq:eqsform}
\end{equation}
with the coefficients as follows: for the commensurate system, 
$\rho_0=25.15\text{ nm}^\inlinesup{-3}$,
$e_0=-6.45\text{ K}$,
$b=17.3\text{ K}$,
$c=15.7\text{ K}$; 
for the system with a vacancy,
$\rho_0=25.51\text{ nm}^\inlinesup{-3}$,
$e_0=-6.34\text{ K}$,
$b=22.5\text{ K}$,
$c=11.7\text{ K}$. 
From the dependence of energy on density, we can extract 
the formation properties according to 
Eqs~(\ref{enthformexpr}--\ref{energyatdensity}).
To accurately calculate the standard deviation 
in the quantities calculated from $e(\rho)$, 
we studied the distribution of each such quantity 
as determined directly by the
statistical errors in the calculation of the energy values.

\begin{figure}
\psfrag{pressurebar}[][c]{$P$, [bar]}
\psfrag{enthalpyandepinK}[][c]{$\Delta E_P \text{ and } h_\text{vac}$, [K]}
\psfrag{melting}{melting}
\includegraphics[angle=0,width=\columnwidth]{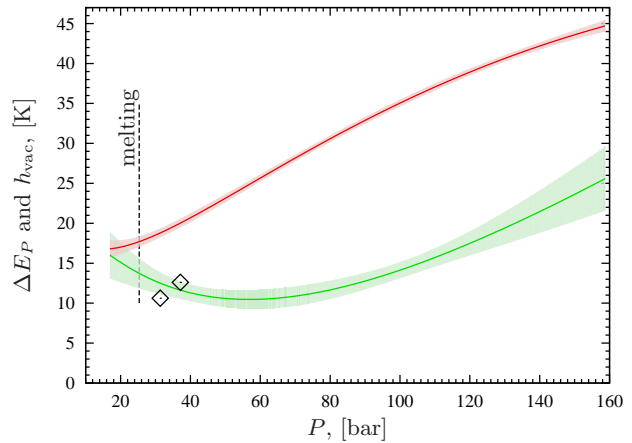}
\caption{\label{fig:epandenthalpy} (Color online) Vacancy formation enthalpy (red, upper curve) 
and isobaric formation energy (green, lower curve).
Shaded areas mark one standard deviation.
Dashed line shows melting pressure.
Diamonds show experimental measurements from Ref.~\cite{Simmons1989} (see text).}
\end{figure}

Vacancy formation energy at fixed density, computed according to Eq.~(\ref{energyatdensity}),
is shown in Fig.~\ref{fig:energyatfixeddensity}. 
At large densities, $\Delta E_\rho$ grows nearly linearly.
The apparent bending at highest considered densities is not 
discernible within the current error levels.
For densities $\rho$ above 30~nm$^\inlinesup{-3}$, 
$\Delta E_\rho$ can be approximated within 0.5~K as $\Delta E_\rho=k (\rho/\rho_r-1)$
with $k=108.2$~K and $\rho_r=25.16$~nm$^\inlinesup{-3}$.
It is notable that $\Delta E_\rho$ approaches a plateau at low densities. 
The lowest value $\Delta E_\rho=16.5\pm1.4\text{ K}$ is reached
at the lowest studied density of $\rho=27.6 \text{ nm}^{\inlinesup{-3}}$.
Fixed-density formation energies obtained by other groups are also plotted for comparison 
in Fig.~\ref{fig:energyatfixeddensity}. 
Our values are higher over a range of densities close to melting.

By matching the pressures, we were able to compute 
the vacancy formation enthalpy and the isobaric formation energy,
shown in Fig.~\ref{fig:epandenthalpy}. 
The experimental results of Fraass et al.~\cite{Simmons1989} are shown 
in the same figure.
Samples in that experiment were grown at constant pressure,
but the experiment itself was reported to be performed at fixed volume. 
The experimental results in Fig.~\ref{fig:epandenthalpy}
are shown at the pressure at which Fraass et al.\ reports the end of 
the solidification process.
The coincidence between the isobaric formation energy and the experimental results compared in this way
calls for more careful examination of the thermodynamic
conditions under which such experiments are carried out.

The minimum that we find in $\Delta E_P$ is unexpected and is not observed in classical solids \cite{Gulseren2003}.
Expanding the equation of state of the system with one vacancy around density $\rho$,
one obtains
\begin{equation*}
\Delta E_P^\text{approx}(P)=\Delta E_\rho(\rho) - (N-1)\beta_\text{inc} P\frac{\Delta P}{\rho}+\mathcal{O}\left(\frac{1}{N}\right)\text{,}
\end{equation*} 
where $\Delta P=P_\text{inc}(\rho)-P_\text{com}(\rho)$ is the 
difference in pressure between the two systems, 
$\beta_\text{inc}$ is the compressibility of the incommensurate system and $P=P_\text{com}(\rho)$.
Using the quantities obtained from our calculations, we find that $\Delta E_P^\text{approx}$
matches the result for $\Delta E_P$ to within 1~K, including the minimum structure.
The minimum in $\Delta E_P$ occurs around the pressure of 57 bar, where the energy 
lowers to $10.5\pm1.2$~K. 
This pressure is coincidentally rather close to the value where Kim and Chan \cite{Chan2006} 
observed a maximum in the supersolid fraction.
Unlike for energy, there is not an  extremum in vacancy formation enthalpy. 
The minimum in energy is balanced by the character
of the dependence of vacancy volume on pressure.
Nonetheless, the second derivative of $h_\text{vac}$ changes sign around this pressure.

\begin{figure}
\psfrag{pressurebar}[][c]{$P$, [bar]}
\psfrag{VVnnc}[][c]{$\Omega_\text{vac}$, [nm$^{3}$]}
\psfrag{VVtoUC}[][c][0.7]{$\Omega_\text{vac}/v_\text{lat}$}
\psfrag{rssor}[][c]{$\rho_\text{s}/\rho$}
\psfrag{melting}{melting}
\psfrag{0}{0}
\psfrag{1}{1}
\includegraphics[angle=0,width=\columnwidth]{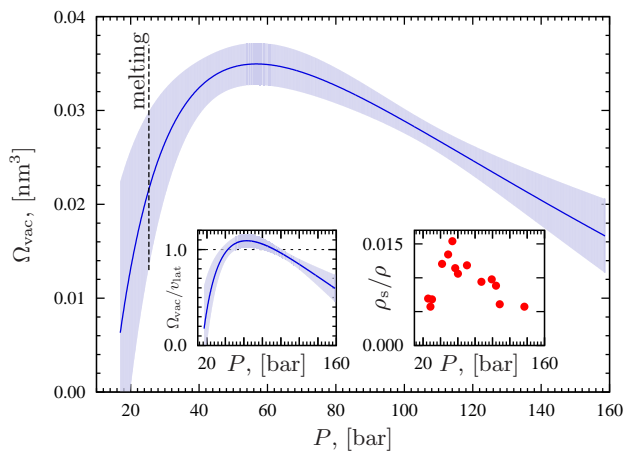}
\caption{\label{fig:vacancyvolume} (Color online) Vacancy formation volume 
as a function of pressure [Eq.~(\ref{vacancyvolume})].
Shaded area indicates statistical error. The curve peaks at $P=57$~bar. 
The left inset shows formation volume of vacancy relative to the volume of the unit cell,
while the inset on the right shows the experimental supersolid fraction from Ref.~\cite{Chan2006}.
}
\end{figure}

The vacancy formation volume $\Omega_\text{vac}$,
shown in Fig.~\ref{fig:vacancyvolume}, is also non-monotonic and 
reaches a maximum at 57 bar, while 
its volume in relation to the unit cell volume of the incommensurate system $v_\text{lat}$
peaks at 63~bar (see left inset in Fig.~\ref{fig:vacancyvolume}; statistical uncertainty does not allow us to distinguish whether
$\Omega_\text{vac}$ indeed peaks above $v_\text{lat}$).
Not only the peak location, but also the overall shape of  $\Omega_\text{vac}(P)$
strikingly resembles the 
pressure dependence of the supersolid fraction as measured by Kim and Chan \cite{Chan2006} 
(shown for comparison on the right inset of Fig.~\ref{fig:vacancyvolume}).
While we do not have a rigorous explanation for this coincidence, it is worth
noting that the peak value of $\Omega_\text{vac}$ is close to $v_\text{lat}$. 
Consider some volume $V=L^\inlinesup{3}$ surrounding a vacancy.
To ensure the correct value of $\Omega_\text{vac}-v_\text{lat}$, the deformation
of lattice lines on the border of this imaginary volume has to scale as $(\Omega_\text{vac}-v_\text{lat})/L^\inlinesup{2}$.
Therefore, the strain on the border of this volume is proportional to $(\Omega_\text{vac}-v_\text{lat})v^\inlinesup{1/3}_\text{lat}/V$.
Volume encompassing a region with the strain above some threshold $\varepsilon_c$ 
is given by $V_c\propto(\Omega_\text{vac}-v_\text{lat})v^\inlinesup{1/3}_\text{lat}/\varepsilon_c$.
This means that the volume of the strained region of the lattice that accompanies a vacant lattice site
in solid helium is strongly reduced at pressures close to 60~bar.

To conclude, we have been able to characterize the isobaric vacancy formation in solid \textsuperscript{4}He.
Thermodynamic vacancy properties were extracted from the calculations of the equations of state of 
solid helium with number of atoms both commensurate and incommensurate with the number of available atomic sites.
Isobaric vacancy formation energy turned out to be significantly lower 
than the energy of formation at fixed density.
The value of the isobaric formation energy is nonetheless 
high enough to exclude the possibility of intrinsic non-interacting 
vacancies in the supersolid experiments.
However, we find a strong resemblance between the pressure 
dependence of vacancy formation volume and experimental results for the supersolid fraction.
The maximum in the formation volume is accompanied by a formation energy minimum at the same pressure.
These coincidences suggest that vacancies may be in fact in some way involved in the supersolidity mechanisms. 

\begin{acknowledgments}
Authors would like to thank Mike Gillan and Moses Chan for helpful discussions.
This work was partially supported by \abbrev{dgi} (Spain) Grant No.\ \abbrev{fis}2008-04403 and
Generalitat de Catalunya Grant No.\ 2009\abbrev{sgr}-1003.
\end{acknowledgments}

%\bibliography{bibs/VacancyPaper}

\end{document}